\documentclass[10pt]{IEEEtran}
\usepackage{hyphenat}
\IEEEoverridecommandlockouts
\usepackage{cite}
\usepackage{amsmath,amssymb,amsfonts}
\usepackage{graphicx}
\usepackage{textcomp}
\usepackage{booktabs}

\usepackage{multirow}
\usepackage{hyperref}
\usepackage{url}
\usepackage{nicefrac}
\usepackage{microtype}
\usepackage{pifont}
\usepackage{xcolor}
\usepackage{stfloats}
\usepackage{algorithm}
\usepackage{orcidlink}
\usepackage{algpseudocode}
\usepackage{tikz}

\usepackage{orcidlink}
\usepackage{comment}
\usepackage{braket}
\usepackage[font=scriptsize,skip=1pt]{caption}
\def\BibTeX{{\rm B\kern-.05em{\sc i\kern-.025em b}\kern-.08em
    T\kern-.1667em\lower.7ex\hbox{E}\kern-.125emX}}
\begin{document}

\title{Quantum Reservoir Computing for Short-Term Power Load Forecasting in Resource-Constrained Energy Systems}

\author{\IEEEauthorblockN{Mansi Od\orcidlink{0009-0005-4618-003X}\textsuperscript{1}, Param Pathak\orcidlink{0009-0003-6419-0915}\textsuperscript{2}, Nouhaila Innan\orcidlink{0000-0002-1014-3457}\textsuperscript{3,4}, and Muhammad Shafique\orcidlink{0000-0002-2607-8135}\textsuperscript{3,4}\\
\IEEEauthorblockA{
\textsuperscript{1}University of Greenwich, London, United Kingdom\\
\textsuperscript{2}Fractal Analytics, \textit{QuantumAI Lab}, Mumbai, Maharashtra, India\\
\textsuperscript{3}eBRAIN Lab, Division of Engineering, New York University Abu Dhabi (NYUAD), Abu Dhabi, UAE\\
\textsuperscript{4}Center for Quantum and Topological Systems (CQTS), NYUAD Research Institute, NYUAD, Abu Dhabi, UAE\\
mansiod2049@gmail.com, param.pathak@fractal.ai, \{nouhaila.innan, muhammad.shafique\}@nyu.edu\\
}}}

\maketitle
\thispagestyle{empty}
\pagestyle{empty}
\begin{abstract}
Short-term load forecasting is essential for reliable energy management, but practical deployment on edge devices requires models that remain accurate under limited memory, finite measurement budgets, and hardware noise. This work proposes a hardware-efficient Quantum Reservoir Computing (QRC) framework for energy load forecasting, where a fixed quantum reservoir transforms temporal input windows into high-dimensional features and only a classical Elastic Net readout is trained. To reduce deployment cost, the trained readout is compressed using post-training fixed-point quantization at bit widths from 8 to 2 bits. The framework is evaluated on the Tetouan and Spain energy load datasets under exact statevector simulation, 512-shot finite sampling, and realistic hardware-noise models from \textit{IBM FakeTorino} and \textit{IBM FakeMarrakesh}. Results show that 6-bit readout precision preserves full-precision forecasting performance while reducing readout memory by 81.2\%. Below this point, degradation becomes dataset dependent, with Tetouan showing stronger sensitivity and Spain degrading more gradually. Hardware-noise validation further shows that the trained readout transfers to noisy reservoir states without retraining. These findings support quantized QRC as a resource-aware forecasting approach for near-term quantum time-series applications.
\end{abstract}


\section{Introduction}

Modern power grids are becoming increasingly distributed, pushing Short-Term Load 
Forecasting~(STLF) from centralized control rooms toward edge nodes embedded in 
smart meters and local controllers~\cite{iqbal2025edge}. Accurate sub-hourly demand 
prediction is indispensable for operational planning, frequency regulation, and 
demand-response dispatch~\cite{millar2025energy}. The volatility and non-stationarity 
of high-frequency consumption data make this a fundamentally difficult sequential 
regression task, and the stakes of forecasting error are high in both economic and 
grid-reliability terms.

Recurrent architectures such as Long Short-Term Memory~(LSTM) networks and Gated Recurrent Units~(GRUs), as well as attention-based Transformers, have set the current performance ceiling for 
STLF~\cite{kong2025deep}. By learning rich temporal dependencies through backpropagation across many layers, these models achieve low forecasting error on standard benchmarks. However, their accuracy comes at a steep hardware cost: millions of trainable parameters, large activation buffers, and repeated floating-point matrix multiplications make inference latency and storage requirements incompatible with the memory and energy limits of resource-constrained edge devices~\cite{mortezanejad2025addressing}.

 Reservoir Computing~(RC) 
offers a structurally different solution to this deployment gap~\cite{habibi2025electrical}. 
Rather than propagating gradients through a recurrent network, RC projects input 
signals into a fixed, randomly initialized dynamical system and 
trains only a linear readout layer on top of the resulting high-dimensional state trajectory. Because the reservoir is never updated, training reduces to a convex readout regression problem, avoiding the costly backpropagation passes required by deep recurrent models. This makes RC attractive for low-power deployment. 
However, classical echo-state reservoirs rely on random recurrent matrices whose 
effective dimensionality is bounded by the number of physical nodes, potentially 
limiting their representational capacity for complex multivariate energy 
signals~\cite{dakheel2025optimizing}.

Quantum Reservoir 
Computing~(QRC) extends the RC paradigm to quantum mechanical systems, where the 
reservoir is a parameterized quantum circuit with frozen weights~\cite{fujii2017harnessing}. 
Even a modest register of $N$ qubits evolves in a Hilbert space of dimension $2^N$, providing access to a high-dimensional nonlinear feature space with a compact physical register.
Superposition and entanglement generate intricate 
correlations across input features that are difficult to replicate in classical 
systems, while the fixed-circuit design preserves the core RC advantage of 
gradient-free training. Critically, QRC sidesteps the barren-plateau problem 
that undermines the trainability of fully parameterized quantum 
models~\cite{mcclean2018barren}, making it a suitable candidate for 
resource-constrained forecasting.

Despite these advantages, a concrete hardware barrier remains. Even when the quantum processing 
stage is fixed, the classical readout layer must store a weight vector whose 
dimensionality scales with the number of measured observables. At full 32-bit 
floating-point precision this readout imposes a non-trivial memory footprint that 
can exceed the on-chip Static Random-Access Memory budget of embedded energy controllers. Post-training 
fixed-point quantization of the readout weights is therefore a necessary step 
toward genuine edge deployability, yet its effect on the rich high-dimensional 
representations produced by a quantum reservoir has not been systematically 
studied. It is unclear whether the informational redundancy embedded in quantum 
feature maps offers any resilience against aggressive low-bit compression, or 
whether quantization noise degrades the forecasting signal in the same way it 
affects classical dense readouts.

\begin{figure}[htpb]
    \centering
    \includegraphics[width=1\linewidth]{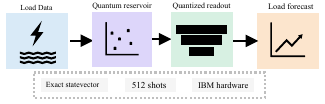}
    \caption{Conceptual overview of the proposed hardware-efficient QRC forecasting framework. Energy load data is encoded into a fixed quantum reservoir, transformed into reservoir features, and mapped to load forecasts through a quantized classical readout. The framework is evaluated under exact simulation, finite-shot sampling, and realistic IBM backend-noise models.}
    \label{intro}
\end{figure}

Through this work, we close this gap by proposing a Hardware-Efficient QRC framework that treats the classical readout as the primary compression target (see Fig. \ref{intro}). 
The quantum reservoir remains fully fixed throughout all experiments; only the 
readout weights are quantized post-training to bit widths 
$k \in \{8, 6, 4, 3, 2\}$. Using the Tetouan and Spain energy load datasets as two distinct forecasting benchmarks, we investigate two core scientific questions:

\begin{itemize}
    \item \textbf{Representation resilience:} Do the high-dimensional quantum 
    feature maps retain sufficient informational structure to support accurate 
    forecasting when the readout is compressed to sub-8-bit precision?
    
    \item \textbf{The accuracy-compactness frontier:} What is the effective 
    accuracy-per-bit trade-off of a QRC readout, and at which bit width does 
    degradation become practically significant?
\end{itemize}

Our results show that 6-bit precision matches full-precision forecasting performance while reducing readout memory by about $81\%$. Below this point, degradation becomes dataset dependent, with stronger sensitivity on the smaller Tetouan setting and smoother degradation on the larger Spain dataset. These findings indicate that QRC readouts can support lightweight forecasting under practical memory constraints.

The rest of the paper is organized as follows:
Sec.~\ref{sec:bg} reviews related work on reservoir computing for load forecasting and quantum-enhanced forecasting on near-term hardware.
Sec.~\ref{sec:methodology}
presents the proposed QRC framework, covering the quantum reservoir 
architecture, temporal kernel aggregation, classical readout training, 
and post-training quantization pipeline.
Sec.~\ref{sec:result} reports 
the results and the discussion. 
Finally, Sec.~\ref{sec:conclusion} concludes with a summary of the main findings  and directions for future work.

\section{Background and Related Work}
\label{sec:bg}
This section reviews prior work along four directions: reservoir computing for load forecasting, quantum-enhanced forecasting on near-term hardware, edge-constrained forecasting, and finite-shot noise in QRC systems. These directions frame the main gap addressed in this work: whether a fixed QRC model can support accurate load forecasting when its readout is compressed to low-bit precision and its reservoir states are evaluated under finite-shot and hardware-noise conditions.


\subsection{Reservoir Computing for Probabilistic Load Forecasting}
The challenge of reconciling forecasting accuracy with computational 
tractability has increasingly motivated the use of Echo State Networks 
(ESNs) in short-term load forecasting. Unlike conventional deep learning 
models that require costly gradient-based optimisation across all layers, 
ESNs fix a randomly initialized sparse recurrent reservoir and train only 
a linear readout, dramatically reducing training overhead. Guerra 
\textit{et al.}~\cite{guerra2023probabilistic} extended this principle by overlaying 
Bayesian uncertainty quantification onto the readout layer, yielding 
calibrated prediction intervals on the ACEA and Spain national grid 
datasets while training significantly faster than autoregressive baselines. 
Melhem \textit{et al.}~\cite{melhem2025edgeai} further proposed a hybrid ESN-GRU 
architecture for real-time edge deployment, achieving an RMSE of 142.3 kW 
while reducing inference latency by 3.2$\times$ and memory footprint by 
47\% on constrained hardware. While these results validate the reservoir computing paradigm for practical smart grid deployment, classical reservoirs remain fundamentally limited in their ability to capture the highly nonlinear demand profiles of modern grids, motivating the exploration of quantum alternatives.

\subsection{Quantum-Enhanced Short-Term Load Forecasting}
Building on this motivation, Habibi \textit{et al.}~\cite{habibi2025electrical} proposed a hybrid quantum-classical 
neural network in which a parameterized quantum circuit functions as a nonlinear feature extractor, achieving an RMSE reduction of approximately 6.4\% over the strongest classical comparator on real power system data while remaining tractable on near-term Noisy Intermediate-Scale Quantum hardware. Pathak \textit{et al.}~\cite{pathak2026late} formalized a QRC framework using Chebyshev encoding and brickwork entanglement, reporting accuracy within 1\% of full-precision baselines on the Tetouan City dataset under finite-shot constraints. However, neither of these works considers the hardware constraints imposed by real edge deployment, particularly the impact of quantized readout and finite measurement budgets.

\subsection{Edge-Constrained Load Forecasting}

Addressing the challenge of establishing accurate forecasting models 
directly on memory-limited edge hardware, Lekidis and Papageorgiou~\cite{lekidis2023edge} proposed a lightweight edge-based energy demand prediction framework. Yet this framework relies entirely on classical models and does not account for the additional noise source inherent to quantum measurement, namely finite-shot sampling.

\subsection{Finite-Shot Noise Analysis in Quantum Reservoir Systems}
This measurement noise represents a fundamental and often overlooked constraint.
Finite measurement budgets impose an irreducible constraint on QRC performance, as evaluating expectation values requires repeated circuit executions, introducing finite-sampling noise even on fault-tolerant hardware~\cite{Kreplin2024reduction}. Ahmed \textit{et al.}~\cite{ahmed2025optimal} show that this noise degrades both recurrent QRC and  Recurrence-Free Quantum Reservoir Computing, affecting the recurrent variant more severely due to noise propagation, a finding validated by Liu \textit{et al.}, who demonstrate that sampling noise undermines multi-step QRC convergence while single-step architectures remain more robust~\cite{liu2026practical}. This motivates reporting results under both exact statevector simulation and 512-shot conditions in the present work ~\cite{hu2024generalization}.

These studies leave open the combined question addressed in this paper: whether a fixed QRC model can retain forecasting accuracy when its readout is compressed to low-bit precision and its reservoir states are evaluated under finite-shot and realistic backend-noise conditions. We address this gap using two energy load forecasting datasets, including zone-level Tetouan demand and national-scale Spain demand.
\begin{figure}[htbp]
    \centering
    \includegraphics[width=1\linewidth]{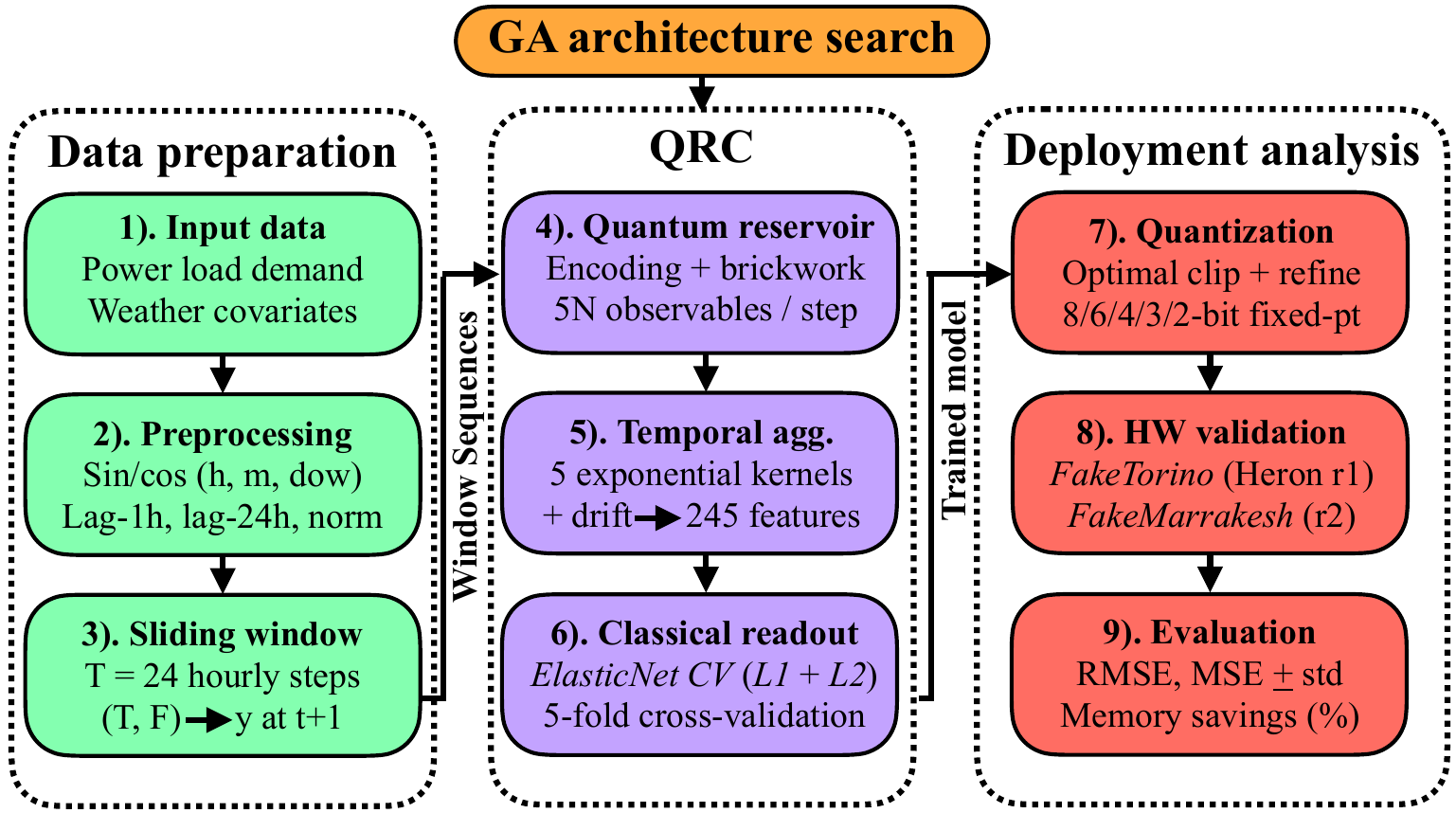}
\caption{Overview of the proposed QRC forecasting framework. The pipeline constructs $T=24$ input windows, extracts $5N$ observables per time step from a fixed quantum reservoir, compresses the temporal states into 245 aggregated features, and trains an Elastic Net readout for load prediction. After training, the readout is quantized to $\{8,6,4,3,2\}$-bit fixed-point precision and evaluated under finite-shot and hardware-noise settings. The GA search is performed offline to select $N$, $L$, encoding, and coupling strength $\kappa$ for each dataset.}
    \label{workflow}
\end{figure}
\section{Methodology}
\label{sec:methodology}
The proposed framework is a hardware-efficient QRC pipeline for short-term load forecasting. Given a multivariate load time series, the model first constructs supervised 24-hour input windows from temporal, meteorological, and autoregressive features. Each window is then processed by a fixed quantum reservoir, where measured observables form a high-dimensional representation of the input dynamics. These reservoir states are compressed through temporal kernel aggregation and mapped to a scalar load forecast using an Elastic Net readout. After training, only the readout weights are quantized to low-bit fixed-point precision and evaluated under finite-shot and hardware-noise settings. A GA-based search is performed offline to select the reservoir configuration for each dataset.

Fig.~\ref{workflow} summarizes the full pipeline, including data preparation, quantum reservoir encoding, temporal aggregation, readout training, post-training quantization, and hardware-noise validation.


\subsection{Data Preparation and Feature Engineering}

The framework assumes access to an hourly power load time series accompanied by meteorological covariates such as temperature, humidity, wind speed, and atmospheric pressure. Additional exogenous variables, for instance cloud cover or electricity price signals, can be included when available. To capture the
periodic structure inherent in energy demand, six cyclical features are constructed using sine and cosine projections of the hour of day, month of year, and day of week. 

This avoids artificial discontinuities between cyclically adjacent values, such as hour 23 and hour 0. Two autoregressive lag features are appended to each sample, specifically the load at one hour prior and at twenty-four hours prior, providing the model with direct access to the most recent intraday pattern. All features are normalized to the unit interval using min-max scaling fit exclusively on the training partition to prevent information leakage. The dataset is split chronologically into 70\% training, 10\% validation, and 20\% test partitions without shuffling, preserving the temporal ordering that is essential for honest evaluation of time series models. A sliding window of $T = 24$ consecutive hourly observations is then applied to construct supervised samples, where each window of shape $(T, F)$ is paired with the scalar load
value at step $t+1$ as the prediction target.

\subsection{Quantum Reservoir Architecture}\label{subsec:qrc_architecture}
\begin{figure}[htbp]
    \centering
    \includegraphics[width=1\linewidth]{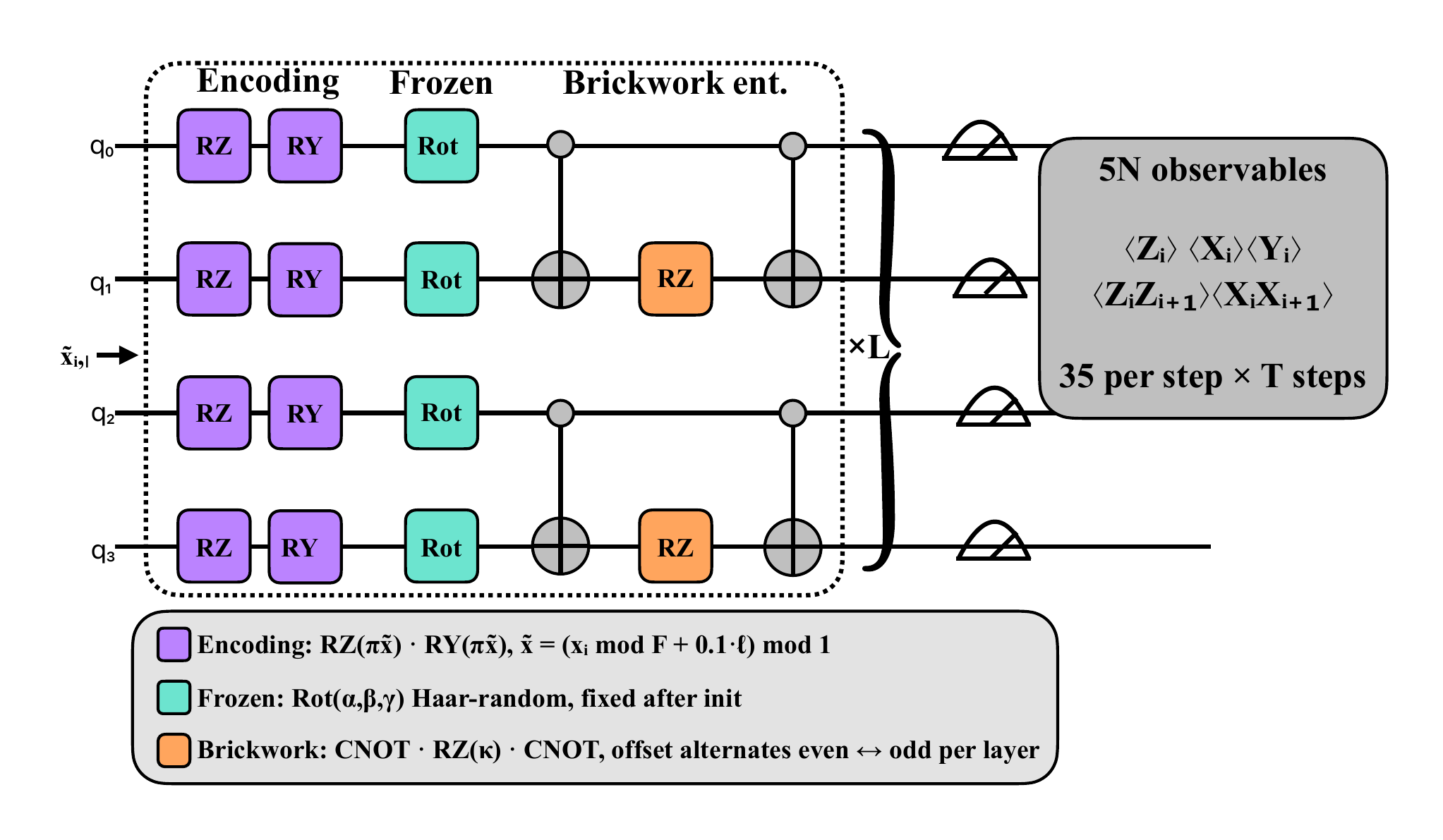}
    \caption{Quantum reservoir circuit at one time step. Each of the $N$ qubits receives the double rotation $R_Z(\pi\tilde{x}) \cdot R_Y(\pi\tilde{x})$ for feature encoding, followed by a frozen Haar random $\mathrm{Rot}(\alpha,\beta,\gamma)$ gate and a brickwork entangling block whose CNOT offset alternates between even and odd pairs across layers. The encode and entangle pattern repeats $L$ times before measurement of single qubit Pauli expectations $\langle Z_i\rangle, \langle X_i\rangle, \langle Y_i\rangle$ and nearest neighbour correlators $\langle Z_iZ_{i+1}\rangle, \langle X_iX_{i+1}\rangle$, yielding $5N$ observables per step.}
    \label{ckt}
\end{figure}

The quantum reservoir is a parameterized circuit whose internal weights are randomly initialized once and then kept fixed, so only the readout layer is trained. This avoids quantum gradient optimization and reduces exposure to barren plateau effects. A genetic algorithm (GA) is used offline to select the optimal reservoir configuration, searching over the number of qubits $N$, the number of entanglement layers $L$, the encoding strategy, and the coupling strength $\kappa$. Fig.~\ref{ckt} illustrates the resulting circuit structure. 

At each time step $t$ within the input window, the normalized feature vector $\mathbf{x}_t \in \mathbb{R}^F$ is encoded into an $N$-qubit register. A layer-dependent shift $\delta_l = 0.1 \cdot l$ is applied to each feature before rotation to break symmetry across layers. The shifted feature for qubit $i$ at layer $l$ is computed as
\begin{equation}
    \tilde{x}_{i,l} = \left(x_{i \bmod F} + \delta_l\right) \bmod 1.
\end{equation}

Each qubit $i$ then receives a double-rotation gate sequence
$R_Z(\pi \tilde{x}_{i,l}) \cdot R_Y(\pi \tilde{x}_{i,l})$, which encodes each feature onto both the azimuthal and polar axes of the Bloch sphere simultaneously. This provides a richer initial state compared to single-axis encodings such as \textit{Chebyshev} angle mapping, which applies only a single $R_Y$ rotation per qubit.

Following the encoding, each qubit receives a general rotation gate $\text{Rot}(\alpha_i, \beta_i, \gamma_i)$ with Haar-randomly sampled angles that remain fixed throughout all experiments. The entanglement structure follows a brickwork topology, where nearest-neighbour CNOT pairs alternate between even and odd offsets across successive layers. Specifically, for layer $l$ the entangling operation is applied to qubit pairs $\{(i, i+1)\}$ with
offset $l \bmod 2$, implementing the pattern
\begin{equation}
    \begin{split}
        \text{CNOT}_{i,i+1} \cdot R_Z(\kappa)_{i+1} \cdot \text{CNOT}_{i,i+1},\\
        i \in \{\text{offset},\ \text{offset}+2,\ \ldots\},
    \end{split}
\end{equation}
where $\kappa$ is the coupling strength selected by the GA. This alternating offset ensures that information propagates across the full qubit register over multiple layers rather than remaining confined to isolated pairs.

After all $L$ layers, the reservoir state is measured. The measurement observable set at each time step consists of single-qubit Pauli expectations and nearest-neighbor two-qubit correlators:
\begin{equation}
\begin{aligned}
    \mathcal{O} ={}& \{Z_i, X_i, Y_i\}_{i=0}^{N-1}
                  \cup \{Z_i Z_{(i+1)\bmod N}\}_{i=0}^{N-1} \\
                  & \cup \{X_i X_{(i+1)\bmod N}\}_{i=0}^{N-1}.
\end{aligned}
\end{equation}

This yields $5N$ expectation values per time step. Over a full input window of $T$ steps, the raw reservoir output consists of $5N \times T$ scalar features, which are then compressed by the temporal aggregation stage described in the following subsection.

\subsection{Temporal Kernel Aggregation}
Rather than passing the full $5N \times T$ state sequence to the readout, temporal kernel aggregation compresses the window into a compact feature vector and reduces overfitting risk. Five exponential decay kernels with
rates $\alpha \in \{0.50, 0.70, 0.85, 0.95, 1.00\}$ are applied across the $T$ time steps. For each kernel rate $\alpha$, the aggregated feature is
\begin{equation}
    \mathbf{f}_\alpha = \sum_{t=0}^{T-1} w_t^\alpha \cdot \mathbf{s}_t,
    \qquad
    w_t^\alpha = \frac{\alpha^{T-1-t}}{\sum_{t'} \alpha^{T-1-t'}},
\end{equation}
where $\mathbf{s}_t \in \mathbb{R}^{5N}$ is the reservoir state at step $t$, and the index $t'$ in the denominator runs over the same range $0$ to $T-1$ as $t$, making the denominator a normalization constant that ensures the kernel weights sum to unity. The five kernels assign different levels of emphasis to
recent versus distant time steps, allowing the readout to access information at multiple temporal scales. The kernel with $\alpha = 0.50$ decays rapidly and focuses almost entirely on the most recent observations, while $\alpha = 1.00$ produces a uniform average that weights all steps equally.

In addition to the five kernel features, two non-weighted summary features are appended. The first is the terminal reservoir state $\mathbf{s}_{T-1}$, which provides direct access to the final measurement output. The second is the first-to-last difference $\mathbf{s}_{T-1} - \mathbf{s}_0$, which captures the net drift of the reservoir dynamics over the input window and encodes how
much the quantum state has evolved during the observation period. The final aggregated feature vector is therefore
\begin{equation}
\begin{aligned}
\mathbf{r} =
\big[&\mathbf{f}_{0.5},\, \mathbf{f}_{0.7},\, \mathbf{f}_{0.85},\,
\mathbf{f}_{0.95},\, \mathbf{f}_{1.0}, \\
&\mathbf{s}_{T-1} - \mathbf{s}_0,\,
\mathbf{s}_{T-1}\big]
\in \mathbb{R}^{(K+2) \times 5N},
\end{aligned}
\end{equation}
where $K = 5$ is the number of decay kernels. This aggregation strategy is inspired by liquid state machine readout methods and retains multi-timescale information while keeping the readout dimensionality tractable for a linear model.

\subsection{Classical Readout Training}

After temporal aggregation, the only trainable component is an Elastic Net readout that maps the compact reservoir vector $\mathbf{r}$ to the scalar load target. This keeps training convex and computationally inexpensive, while the $\ell_1$ term enables sparse selection over reservoir observables and the $\ell_2$ term improves numerical stability.

\begin{figure}[htbp]
    \centering
    \includegraphics[width=1\linewidth]{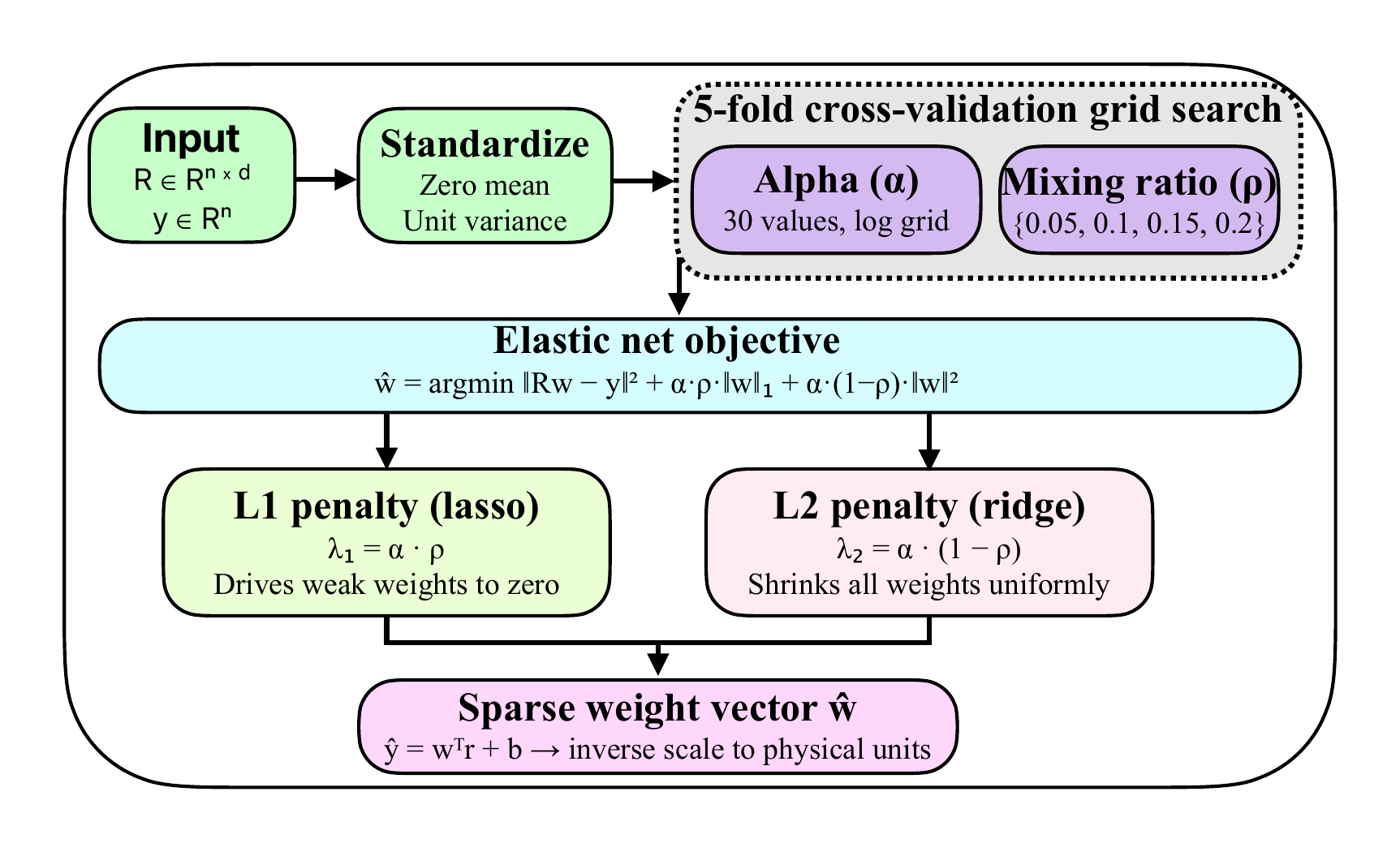}
    \vspace{-0.8cm}
    \caption{ElasticNet readout training pipeline. Standardized reservoir features $\mathbf{R}$ and targets $\mathbf{y}$ feed a 5 fold cross validated grid search over the regularization strength $\alpha$ (30 log spaced values) and the mixing ratio $\rho \in \{0.05, 0.1, 0.15, 0.2\}$, which set the $\ell_1$ and $\ell_2$ penalty weights $\lambda_1 = \alpha\rho$ and $\lambda_2 = \alpha(1-\rho)$. The selected combination yields a sparse weight vector $\hat{\mathbf{w}}$ whose predictions are inverse scaled to physical units.}
    \label{CV}
\end{figure}

The aggregated reservoir features are first standardized to zero mean and unit variance using a scaler fitted exclusively on the training partition. The readout is then trained using Elastic Net regression, which combines $\ell_1$ and $\ell_2$ regularization into a single objective. Fig.~\ref{CV} illustrates this training pipeline. The optimization problem solved by the readout is
\begin{equation}
    \hat{\mathbf{w}} = \arg\min_{\mathbf{w}}
    \left\| \mathbf{R}\mathbf{w} - \mathbf{y} \right\|_2^2
    + \lambda_1 \left\| \mathbf{w} \right\|_1
    + \lambda_2 \left\| \mathbf{w} \right\|_2^2,
\end{equation}
where $\mathbf{R} \in \mathbb{R}^{n \times d}$ is the matrix of aggregated reservoir states over all $n$ training samples, $\mathbf{y}$ is the normalized target vector, and $d = (K+2) \times 5N$ is the readout dimensionality. The
two penalty terms are controlled by a shared regularization strength $\alpha$ and a mixing parameter $\rho$ through the relationships $\lambda_1 = \alpha \cdot \rho$ and $\lambda_2 = \alpha \cdot (1 - \rho)$. The $\ell_1$ component promotes sparsity in the weight vector by driving small or irrelevant coefficients exactly to zero, which acts as an automatic feature selection mechanism over the reservoir observables. The $\ell_2$ component shrinks the remaining nonzero weights toward zero, preventing any single observable from dominating the prediction and improving numerical stability.

The optimal values of $\alpha$ and $\rho$ are selected jointly through five-fold cross-validation. The regularization strength $\alpha$ is searched over a logarithmic grid of 30 candidate values, while the mixing parameter $\rho$ is evaluated over $\{0.05, 0.1, 0.15, 0.2\}$. This range of $\rho$ values keeps the readout in a regime where moderate sparsity is encouraged without aggressively zeroing out too many features. The combination that achieves the lowest cross-validated error on the training partition is selected as the final model. The prediction for an unseen sample is then computed as
\begin{equation}
    \hat{y} = \mathbf{w}^\top \mathbf{r} + b,
\end{equation}
where $b$ is the learned bias term, and the output is inverse-transformed through the min-max scaler to recover predictions in the original physical units of the target variable.

\subsection{Post-Training Quantization}

\begin{figure}[htbp]
    \centering
    \includegraphics[width=1\linewidth]{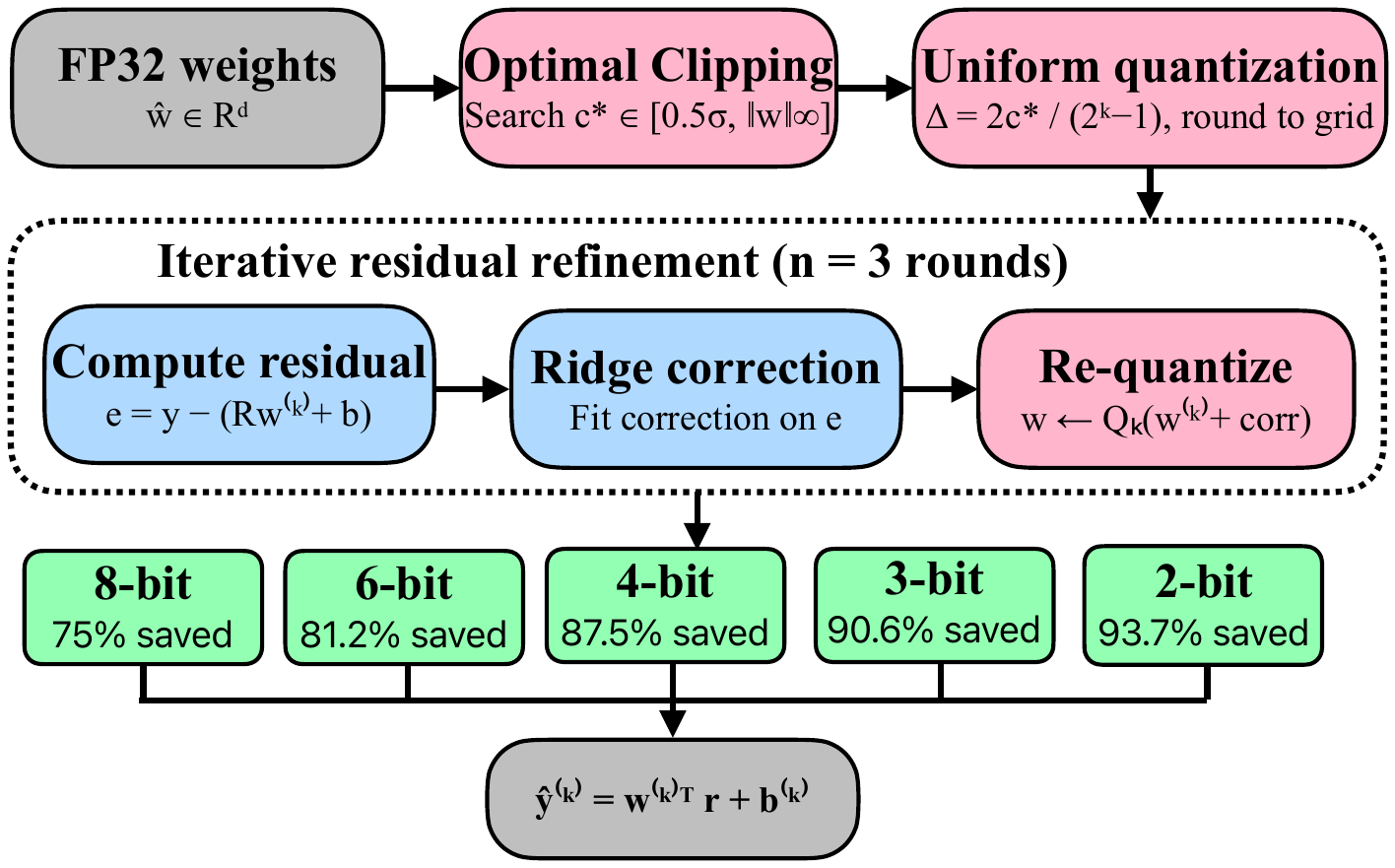}
    \caption{Post training quantization (PTQ) pipeline. The trained FP32 weights pass through an optimal clipping search over $[0.5\sigma, \|\mathbf{w}\|_\infty]$ followed by uniform rounding to a $k$ bit grid with step size $\Delta = 2c^*/(2^k-1)$. Three rounds of iterative residual refinement then fit a Ridge correction on the prediction error and re-quantize, suppressing the systematic bias that grows at low bit widths. The procedure produces five deployable readouts at $\{8,6,4,3,2\}$ bits with corresponding memory savings of $\{75, 81.2, 87.5, 90.6, 93.7\}\%$.}
    \label{qtzn}
\end{figure}

A key advantage of confining all trainable parameters to a single linear readout is that the resulting weight vector can be compressed through post-training quantization without modifying the quantum reservoir or repeating
any quantum computation. This subsection describes the procedure used to reduce the readout from 32-bit floating point to low-bit fixed-point representations,
targeting deployment on resource-constrained edge hardware where memory and arithmetic precision are limited. Fig.~\ref{qtzn} illustrates the full quantization pipeline.

\label{subsec:post-training-qtzn}

For a given target bit width $k \in \{8, 6, 4, 3, 2\}$, the first step is to determine an optimal clipping threshold $c^*$ that defines the representable range of the weight vector. Values outside this range are saturated to the boundary. The threshold is found by grid search over the interval $[0.5\sigma,\, \|\mathbf{w}\|_\infty]$, where $\sigma$ is the standard deviation of the trained weights and $\|\mathbf{w}\|_\infty$ is the largest absolute weight value. The objective of the search is to minimize the mean squared error between the original and quantized weights:
\begin{equation}
    c^* = \arg\min_{c} \, \mathbb{E}\!\left[\left(w - Q_k(w, c)\right)^2\right],
\end{equation}
where $Q_k(w, c)$ is the uniform quantization function defined as $Q_k(w, c) = \Delta \cdot \text{round}(\text{clip}(w, -c, c) / \Delta)$ and $\Delta = 2c / (2^k - 1)$ is the quantization step size, which determines the spacing between adjacent representable values at bit width $k$.

Once the clipping threshold is fixed, the weights are quantized to the nearest grid point. However, direct quantization can introduce systematic bias in the predictions, particularly at aggressive bit widths such as 3-bit or 2-bit. To mitigate this, an iterative residual refinement procedure is applied over three correction rounds. In each round, the residual between the true training targets and the predictions of the current quantized model is computed. A Ridge regression model is then fitted on this residual to produce a correction term,
and the corrected weights are re-quantized. This process progressively reduces the quantization error by allowing the model to compensate for the distortion introduced by rounding. The final quantized prediction for an unseen sample is
\begin{equation}
    \hat{y}^{(k)} = \left(\mathbf{w}^{(k)}\right)^\top \mathbf{r} + b^{(k)},
\end{equation}
where $\mathbf{w}^{(k)}$ and $b^{(k)}$ are the refined quantized weight vector and bias at bit width $k$, and $\mathbf{r}$ is the aggregated reservoir feature vector. Memory savings relative to the original 32-bit representation are computed as $(1 - k/32) \times 100\%$, yielding reductions of 75\%, 81.25\%, 87.5\%, 90.6\%, and 93.75\% for 8, 6, 4, 3, and 2-bit representations respectively. 

Algorithm~\ref{alg:qrc_pipeline} summarizes the full training and quantized deployment procedure. At inference time, the fixed reservoir encoding, temporal aggregation, and quantized readout forward pass are executed using the stored low-bit weights; no quantum gradient computation is required.

\begin{algorithm}
\caption{QRC Pipeline with Post-Training Quantization}
\footnotesize
\label{alg:qrc_pipeline}
\begin{algorithmic}[1]
\Require Multivariate time series $\mathcal{D} = \{(\mathbf{x}_t, y_t)\}_{t=1}^{N}$, target bit widths $\mathcal{B} = \{8, 6, 4, 3, 2\}$, window length $T$, number of qubits $N_q$, number of layers $L$, coupling strength $\kappa$, refinement rounds $n_{\text{iter}}$
\Ensure Load forecasts $\hat{y}^{(k)}$ and memory savings for each $k \in \mathcal{B}$

\Statex \textbf{// Stage 1: Data Preparation}
\State Construct cyclical time features (hour, month, day of week) and autoregressive lags ($y_{t-1}$, $y_{t-24}$)
\State Normalize all features to $[0, 1]$ using training partition statistics
\State Partition chronologically into train, validation, and test sets
\State Form sliding window samples $(\mathbf{X}_i, y_i)$ of shape $(T, F)$

\Statex \textbf{// Stage 2: Quantum Reservoir Encoding}
\State Initialize $N_q \times L$ frozen rotation angles via Haar random sampling
\For{each window sample $\mathbf{X}_i$}
    \For{each time step $t = 1, \ldots, T$}
        \State Encode features into $N_q$ qubits using layer shifted double rotation gates
        \State Apply frozen brickwork entanglement with coupling $\kappa$
        \State Measure $5N_q$ observables (single qubit Pauli and nearest neighbour correlators)
    \EndFor
    \State Aggregate $T$ step measurements into compact vector $\mathbf{r}_i$ using $K{=}5$ exponential decay kernels, terminal state, and drift term
\EndFor

\Statex \textbf{// Stage 3: Classical Readout Training}
\State Standardize reservoir features using training set statistics
\State Train Elastic Net readout via cross validated grid search over regularization strength and sparsity ratio
\State Obtain weight vector $\hat{\mathbf{w}}$ and bias $b$
\State Compute baseline predictions $\hat{y}_i = \hat{\mathbf{w}}^\top \mathbf{r}_i + b$ and invert scaling

\Statex \textbf{// Stage 4: Post-Training Quantization}
\For{each bit width $k \in \mathcal{B}$}
    \State Search for optimal clipping threshold $c^*$ minimizing quantization distortion
    \State Quantize weights to $k$ bit fixed point grid with step size $\Delta = 2c^* / (2^k - 1)$
    \For{round $= 1, \ldots, n_{\text{iter}}$}
        \State Compute prediction residual under current quantized weights
        \State Fit ridge correction on residual and update weights
        \State Re-quantize corrected weights to $k$ bit grid
    \EndFor
    \State Store quantized forecasts $\hat{y}^{(k)}$ and memory saving $(1 - k/32) \times 100\%$
\EndFor

\State \Return $\{\hat{y}^{(k)}, \text{memory saving}^{(k)}\}$ for all $k \in \mathcal{B}$
\end{algorithmic}
\end{algorithm}

The four stages in the algorithm map directly to the architectural blocks in fig.~\ref{workflow}. Stage 1 corresponds to the data preparation block, Stage 2 to the quantum reservoir and temporal aggregation blocks and, Stage 3 to the classical readout block, and Stage 4 to the quantization and deployment analysis block. At inference time, only Stages 2 and the forward pass of Stage 3 are executed, since the trained and quantized weights are stored on device. The quantum reservoir parameters remain frozen throughout, meaning no quantum gradient computation is performed at any point during training or deployment.

\subsection{Evaluation Protocol and Hardware Noise Simulation}

All experiments are conducted across two independent random seeds to account for variability in the stochastic initialization of reservoir weights. Each seed generates a different set of Haar random angles for the frozen rotation gates, producing a distinct reservoir instance that processes the same input through a different unitary evolution. Reporting results as the mean and standard deviation across seeds ensures that the observed performance reflects the general behavior of the framework rather than a single favorable configuration.

Each seed configuration is further evaluated under two measurement settings. The first setting uses exact statevector simulation, denoted as $\text{shots} = \text{None}$, where expectation values are computed analytically without any sampling noise. This serves as the noiseless reference and represents the best achievable accuracy for a given reservoir. The second setting uses 512 shots per circuit execution, which introduces finite sampling noise into the expectation value estimates. This emulates the statistical fluctuations that a physical quantum processor would produce when operating with a limited measurement budget. Comparing these two settings isolates the effect of shot noise on forecasting accuracy, independent of any device level imperfections.

Forecasting quality is assessed using a suite of metrics selected for their relevance to the energy load forecasting community. RMSE and MAE measure prediction accuracy in the native physical units of each dataset:
\begin{equation}
\text{RMSE} = \sqrt{\frac{1}{n}\sum_{i=1}^{n}(\hat{y}_i - y_i)^2}, \quad
\text{MAE} = \frac{1}{n}\sum_{i=1}^{n}|\hat{y}_i - y_i|,
\end{equation}
where $\hat{y}_i$ and $y_i$ are the predicted and actual load values for the $i$\textsuperscript{th} test sample and $n$ is the total number of test samples. To enable comparison across datasets with different scales, the mean absolute percentage error (MAPE) expresses the average prediction deviation as a fraction of the true load:
\begin{equation}
\text{MAPE} = \frac{100}{n}\sum_{i=1}^{n}\left|\frac{\hat{y}_i - y_i}{y_i}\right| \%.
\end{equation}

The coefficient of determination $R^2$ quantifies the proportion of variance in the observed load that the model explains, with a value of 1 indicating a perfect fit and values near 0 indicating performance no better than predicting the mean:
\begin{equation}
R^2 = 1 - \frac{\sum_{i=1}^{n}(\hat{y}_i - y_i)^2}{\sum_{i=1}^{n}(y_i - \bar{y})^2},
\end{equation}
where $\bar{y}$ is the mean of the observed values. Following ASHRAE Guideline~14, which is widely adopted for validating energy models in building and grid applications, two additional metrics are reported. The coefficient of variation of RMSE (CVRMSE) normalizes the root mean squared error by the mean observed load, providing a scale independent measure of prediction spread:
\begin{equation}
\text{CVRMSE} = \frac{\text{RMSE}}{\bar{y}} \times 100\%,
\end{equation}

The normalized mean bias error (NMBE) captures systematic directional tendency in the predictions, indicating whether the model consistently overpredicts or underpredicts demand:
\begin{equation}
\text{NMBE} = \frac{\sum_{i=1}^{n}(\hat{y}_i - y_i)}{n \cdot \bar{y}} \times 100\%.
\end{equation}

Finally, because grid operators are most sensitive to forecasting errors during periods of high demand, when reserve margins are thin and supply shortfalls carry the greatest operational risk, a peak load MAPE is computed over the top 10\% of hours ranked by actual load. This metric isolates model reliability precisely where accurate forecasting matters most for grid stability.

Beyond noiseless and finite shot simulation, the framework is validated under realistic hardware noise to assess whether a model trained under ideal conditions can generalize to the imperfections of physical quantum processors. Two \textit{IBM} fake backends are used for this purpose. \textit{FakeTorino} is a calibrated noise model of the \textit{IBM Torino} processor based on the \textit{Heron~r1} architecture with 133 qubits. \textit{FakeMarrakesh} is a calibrated noise model of the \textit{IBM Marrakesh} processor based on the \textit{Heron~r2} architecture with 156 qubits. Both backends incorporate gate error rates, readout errors, and decoherence parameters extracted from real hardware calibration data, providing a faithful approximation of the noise environment on current superconducting quantum devices. For these experiments, inference is performed on a subset of test samples using 1024 shots per circuit. The saved readout model from the noiseless training phase is loaded and applied directly to the noisy reservoir states without any retraining or fine tuning. This inference only protocol isolates the effect of hardware noise on the quantum reservoir output from the readout optimization, testing whether the framework remains operationally viable under realistic deployment conditions.

\section{Results and Discussion}
\label{sec:result}
\subsection{Experimental Setup}
The proposed framework is evaluated on two publicly available energy load forecasting datasets that differ in spatial scale, sampling resolution, and feature composition. The Tetouan dataset~\cite{salam2018tetouan} records power consumption from three distribution zones in Tetouan, Morocco, at ten-minute intervals over approximately one year. Zone~1 is used as the prediction target in kilowatt hours (kWh). After hourly resampling and the preprocessing described in Sec.~\ref{sec:methodology}, the final dataset contains approximately 8,400 samples with 11 input features.

The Spain dataset~\cite{jhana2019spain} contains four years of hourly national grid records, with total load demand in megawatts (MW) as the prediction target. Energy records are merged with Madrid weather data using UTC-aligned timestamps and then prepared using the same feature construction, scaling, chronological split, and sliding-window protocol. The final dataset contains approximately 35,000 samples with 13 input features.



All quantum circuit simulations are implemented in \textit{PennyLane} using the \texttt{lightning.qubit} backend and executed on an \textit{NVIDIA A100 GPU} through \textit{Google Colab Pro+}. The classical readout is trained using Elastic Net regression from scikit-learn. The regularization strength is selected by five-fold cross-validation over a logarithmic grid of 30 candidates, while the mixing parameter is searched over $\{0.05, 0.1, 0.15, 0.2\}$. Each configuration is evaluated over two random seeds, $\{42,7\}$, under exact statevector simulation and finite-shot simulation with 512 shots per circuit.
PTQ is applied to the trained readout weights at bit widths $\{8,6,4,3,2\}$ using three rounds of iterative residual refinement. Hardware-noise validation is performed using the \textit{IBM FakeTorino} and \textit{IBM FakeMarrakesh} backends through the \texttt{qiskit.remote} \textit{PennyLane} device with 1024 shots per circuit. The trained readout models and feature scalers are serialized using joblib so that the hardware-noise experiments perform inference only, without retraining.
\subsection{Selected Reservoir Architectures}

A GA-based search is used to select the reservoir configuration for each dataset, following the design space defined in Sec.~\ref{subsec:qrc_architecture}. The search considers the number of qubits, entanglement depth, encoding strategy, and coupling strength. In the experiments, the GA uses a population of six individuals over three generations, with validation RMSE used as the fitness criterion. Table~\ref{tab:arch_comparison} reports the selected architectures.
\begin{table}[htpb]
\centering
\caption{Reservoir architectures selected for each dataset.}
\label{tab:arch_comparison}
\begin{tabular}{lcc}
\toprule
\textbf{Parameter} & \textbf{Tetouan} & \textbf{Spain} \\
\midrule
Qubits ($N$) & 7 & 7 \\
Layers ($L$) & 4 & 3 \\
Encoding & Chebyshev & $R_Z$--$R_Y$ \\
Coupling ($\kappa$) & $\pi/2$ & $\pi/3$ \\
Window ($T$) & 24 & 24 \\
Input features & 11 & 13 \\
Observables per step & 35 & 35 \\
Aggregated features & 245 & 245 \\
$\ell_1$ ratio ($\rho$) & 0.1 & 0.1 \\
Val RMSE & 3142.28 & 3706.30 \\
\bottomrule
\end{tabular}
\end{table}

Both datasets select seven qubits and a 24-hour window, yielding 35 expectation values per time step and 245 aggregated features after temporal kernel compression. This common feature dimensionality suggests that the same reservoir size is sufficient for both forecasting tasks, despite differences in dataset scale and feature composition.

The dataset-specific differences appear in the encoding and reservoir dynamics. Tetouan selects Chebyshev encoding with four entanglement layers and stronger coupling, $\kappa=\pi/2$. Spain selects $R_Z$--$R_Y$ double rotation with three layers and weaker coupling, $\kappa=\pi/3$. This indicates that Tetouan benefits from a deeper and more strongly coupled reservoir, while Spain favors richer input rotations with a shallower circuit. Since Spain includes a broader set of exogenous variables, the dual-axis encoding can provide a more expressive initial state without requiring additional entanglement depth.
\begin{table*}[htpb]  
\centering
\caption{Quantization performance across bit widths for both datasets. Values are averaged over two random seeds. Standard deviations remain below 1\% of the mean for all configurations at 6 bits and above.}
\label{tab:quantization_combined}
\footnotesize
\setlength{\tabcolsep}{4pt}
\begin{tabular}{cl c ccccc ccccc}
\toprule
& & & \multicolumn{5}{c}{\textbf{Tetouan (kWh)}} & \multicolumn{5}{c}{\textbf{Spain (MW)}} \\
\cmidrule(lr){4-8} \cmidrule(lr){9-13}
\textbf{Setting} & \textbf{Bits} & \textbf{Mem\,(\%)} & \textbf{RMSE} & \textbf{MAPE} & $\mathbf{R^2}$ & \textbf{CVRMSE} & \textbf{Pk\,MAPE} & \textbf{RMSE} & \textbf{MAPE} & $\mathbf{R^2}$ & \textbf{CVRMSE} & \textbf{Pk\,MAPE} \\
\midrule
\multirow{6}{*}{\rotatebox{90}{Noiseless}}
 & FP32 & ---   & 3359 & 9.81  & 0.695 & 11.42 & 6.39  & 3084 & 8.56  & 0.535 & 10.66 & 13.21 \\
 & 8    & 75.0  & 3355 & 9.81  & 0.696 & 11.41 & 6.32  & 3084 & 8.56  & 0.535 & 10.67 & 13.19 \\
 & \textbf{6}    & \textbf{81.2}  & \textbf{3348} & \textbf{9.77}  & \textbf{0.697} & \textbf{11.38} & \textbf{6.42}  & \textbf{3090} & \textbf{8.57}  & \textbf{0.533} & \textbf{10.68} & \textbf{13.21} \\
 & 4    & 87.5  & 3707 & 11.12 & 0.622 & 12.60 & 5.63  & 3150 & 8.73  & 0.515 & 10.89 & 13.07 \\
 & 3    & 90.6  & 3900 & 12.04 & 0.588 & 13.26 & 5.44  & 3217 & 8.94  & 0.494 & 11.12 & 14.08 \\
 & 2    & 93.8  & 4216 & 13.24 & 0.500 & 14.33 & 5.65  & 3387 & 9.57  & 0.437 & 11.71 & 14.71 \\
\midrule
\multirow{6}{*}{\rotatebox{90}{512 shots}}
 & FP32 & ---   & 3352 & 9.58  & 0.697 & 11.40 & 7.13  & 3083 & 8.55  & 0.535 & 10.66 & 13.12 \\
 & 8    & 75.0  & 3347 & 9.54  & 0.697 & 11.38 & 7.14  & 3083 & 8.55  & 0.535 & 10.66 & 13.09 \\
 & \textbf{6}    & \textbf{81.2}  & \textbf{3286} & \textbf{9.38}  & \textbf{0.708} & \textbf{11.17} & \textbf{7.02}  & \textbf{3080} & \textbf{8.52}  & \textbf{0.536} & \textbf{10.65} & \textbf{13.09} \\
 & 4    & 87.5  & 4009 & 11.85 & 0.554 & 13.63 & 6.52  & 3116 & 8.66 & 0.525 & 10.78 & 13.04 \\
 & 3    & 90.6  & 5073 & 15.95 & 0.274 & 17.25 & 8.27  & 3222 & 8.94 & 0.492 & 11.14 & 14.04 \\
 & 2    & 93.8  & 5535 & 15.64 & 0.153 & 18.82 & 6.15  & 3406 & 9.64 & 0.430 & 11.78 & 13.61 \\
\bottomrule
\end{tabular}
\end{table*}
\subsection{Quantization Performance and Memory Reduction}
\label{subsec:quantization_performance}

The first objective of the evaluation is to determine how far the QRC readout can be compressed before forecasting accuracy begins to degrade. Since the quantum reservoir remains fixed, quantization affects only the trained Elastic Net readout weights. This makes the analysis a direct test of whether the reservoir features contain enough redundancy to tolerate low-bit fixed-point deployment. Table~\ref{tab:quantization_combined} reports the results across five quantization levels, two datasets, and two simulation regimes.

A consistent pattern appears across both datasets: the readout remains stable down to 6-bit precision. In the noiseless setting, the 6-bit model closely matches the FP32 baseline on both Tetouan and Spain, with only minor changes in RMSE, MAPE, and $R^2$. Under 512-shot simulation, the same trend is preserved, and 6-bit quantization even produces a small improvement in several metrics. This suggests that moderate rounding of the readout weights does not disrupt the forecasting function learned from the reservoir features.

\begin{figure*}[htpb]
    \centering
    \includegraphics[width=1\linewidth]{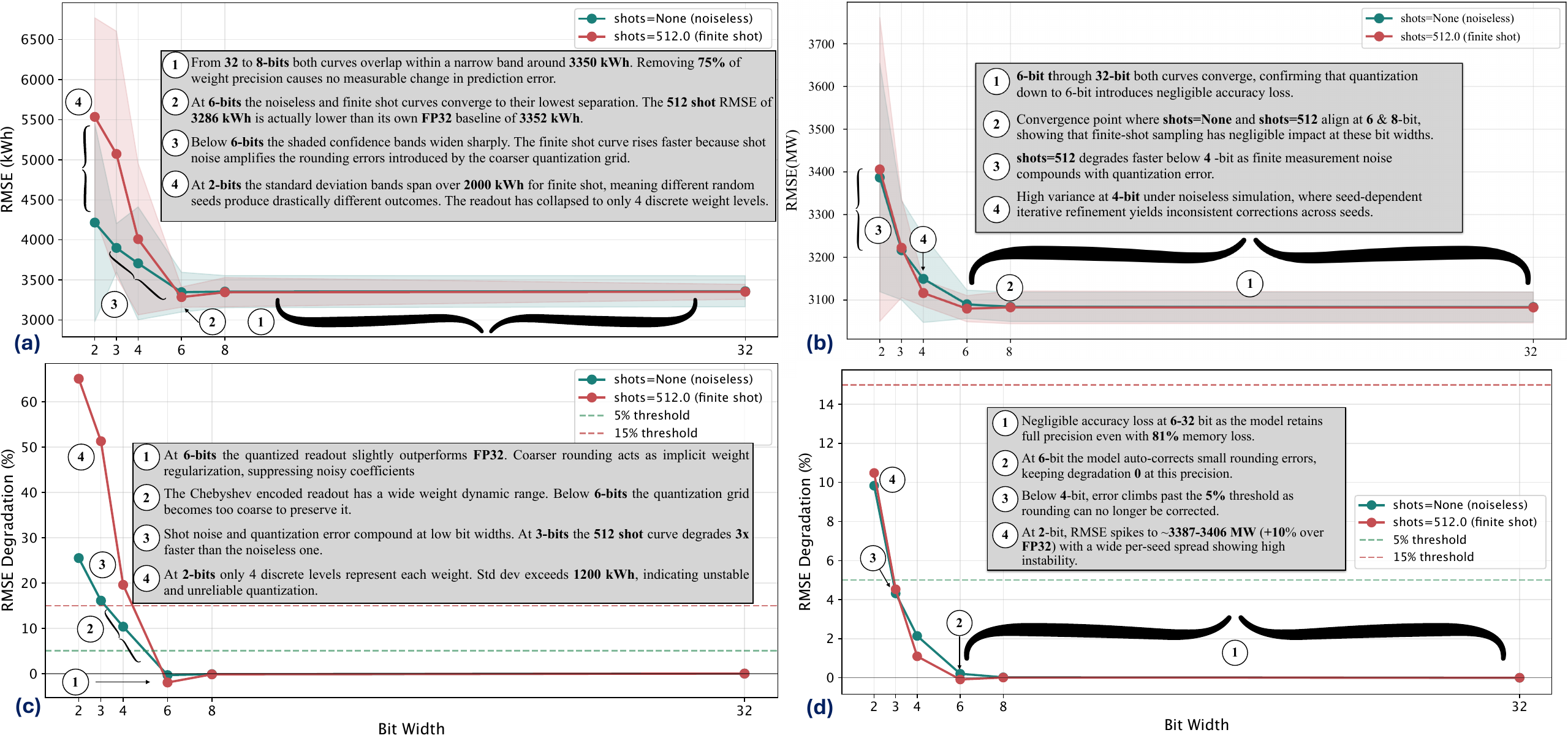}
    \caption{Bit-width sensitivity of the QRC readout for Tetouan and Spain under noiseless and 512-shot simulation. Panels (a) and (c) show the Tetouan results, while panels (b) and (d) show the Spain results. For both datasets, the readout remains stable from FP32 down to 6-bit precision, corresponding to an 81.2\% memory reduction. Below 6 bits, the degradation becomes more visible, with Tetouan showing stronger sensitivity under finite-shot simulation and Spain degrading more gradually.}
    \label{fig:bitwidth_sensitivity}
\end{figure*}
Fig.~\ref{fig:bitwidth_sensitivity} further clarifies the dataset-level differences below 6 bits. Tetouan is more sensitive to aggressive compression: under 512-shot simulation, RMSE increases from 3286~kWh at 6 bits to 4009~kWh at 4 bits, then to 5073~kWh and 5535~kWh at 3 and 2 bits, respectively. Spain degrades more gradually, with RMSE increasing from 3080~MW at 6 bits to 3116~MW at 4 bits, then to 3222~MW and 3406~MW at 3 and 2 bits.

These results identify 6-bit precision as the most reliable operating point. It reduces readout memory by 81.2\% while preserving full-precision performance across datasets and simulation regimes. Lower bit widths provide additional memory savings, but the resulting accuracy loss becomes dataset dependent and is especially pronounced under finite-shot sampling. Therefore, the 6-bit readout offers the best accuracy-memory trade-off for the proposed QRC framework.

\subsection{Forecasting Behavior at the 6-bit Operating Point}
\begin{figure*}[htpb]
    \centering
    \includegraphics[width=1\linewidth]{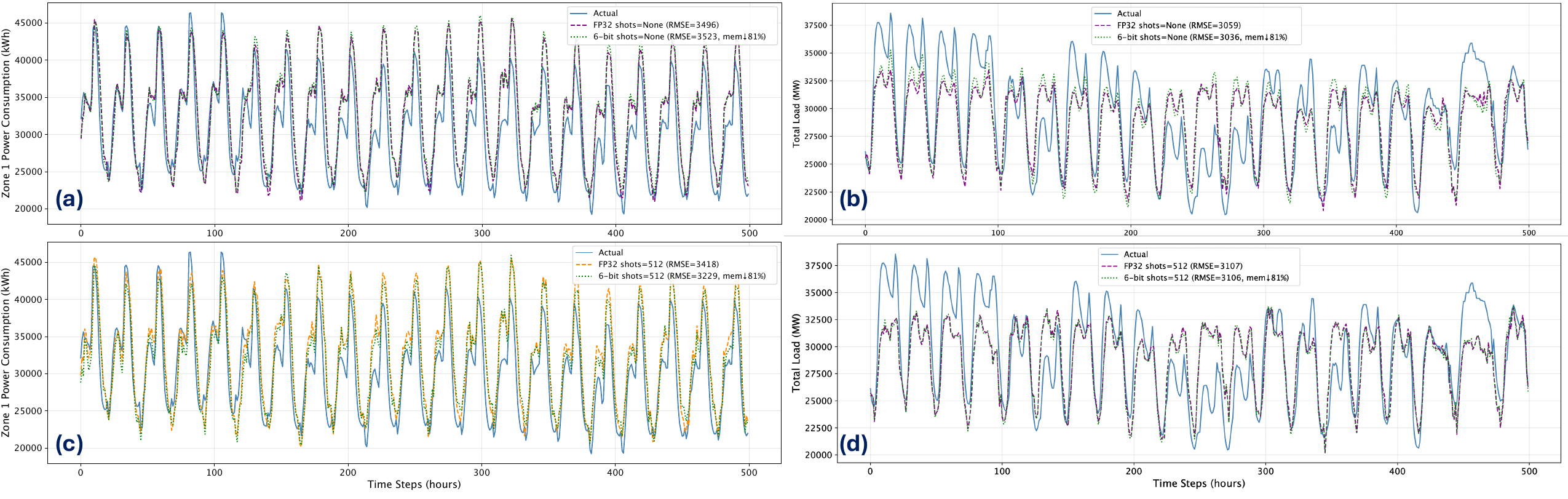}
    \caption{Comparison of FP32 and 6-bit QRC forecasts over 500-hour test windows for the Tetouan and Spain datasets under noiseless and 512-shot simulation (for one fixed seed). Panels (a) and (c) correspond to Tetouan, while panels (b) and (d) correspond to Spain. In both datasets, the 6-bit readout closely follows the FP32 baseline under both simulation settings, showing that the temporal forecast structure is preserved after quantization. The agreement remains strong under finite-shot simulation, confirming that 6-bit quantization provides substantial memory reduction with minimal effect on prediction quality.}
    \label{fig:forecast_6bit_comparison}
\end{figure*}

For Tetouan, the FP32 and 6-bit forecasts remain closely aligned in both settings. Under noiseless simulation, RMSE changes from 3496 to 3523~kWh, corresponding to a difference of only 27~kWh across the 500-hour window. Under 512-shot simulation, the gap increases from 3286 to 3391~kWh, but the quantized curve still follows the full-precision trajectory without visible drift. This indicates that finite-shot sampling slightly increases the numerical gap but does not distort the forecast pattern.
For Spain, the effect of quantization is even smaller. Under noiseless simulation, RMSE changes from 3109 to 3113~MW, corresponding to a $+0.13\%$ penalty. Under 512-shot simulation, the 6-bit readout slightly improves RMSE from 3107 to 3106~MW. The quantized model therefore preserves the main daily load behavior while reducing readout memory by about 81\%.

These results show that the 6-bit readout preserves the forecast trajectory in both datasets, not only the aggregate performance metrics. This is important for load forecasting because the temporal profile of the prediction, especially around transitions and peak-demand intervals, is as important as the average error.

\subsection{Hardware Noise Validation}
\begin{table*}[htpbt]
\centering
\caption{Hardware noise validation on a 50 sample test slice using 1024 shots per circuit. The simulation baseline uses \texttt{lightning.qubit} with matched shot count. Negative degradation indicates that hardware noise improved accuracy.}
\label{tab:hardware_combined}
\footnotesize
\setlength{\tabcolsep}{4pt}
\begin{tabular}{ll l ccccc ccccc}
\toprule
& & & \multicolumn{5}{c}{\textbf{Tetouan (kWh)}} & \multicolumn{5}{c}{\textbf{Spain (MW)}} \\
\cmidrule(lr){4-8} \cmidrule(lr){9-13}
\textbf{Backend} & \textbf{Corr.} & \textbf{Source} & \textbf{RMSE} & \textbf{MAPE} & $\mathbf{R^2}$ & \textbf{CVRMSE} & \textbf{Pk\,MAPE} & \textbf{RMSE} & \textbf{MAPE} & $\mathbf{R^2}$ & \textbf{CVRMSE} & \textbf{Pk\,MAPE} \\
\midrule
\textit{FakeTorino}      & 0.969 & Simulation  & 2172 & 5.00 & 0.882 & 6.63 & 4.19 & 3736 & 9.24 & 0.373 & 11.44 & 15.62 \\
\textit{(Heron r1,}      &       & Hardware    & \textbf{2110} & \textbf{5.06} & \textbf{0.889} & \textbf{6.44} & \textbf{4.67} & 4048 & 10.03 & 0.264 & 12.40 & 16.35 \\
\textit{133 qubits)}     &       & Degradation & \multicolumn{5}{c}{$-$2.9\%} & \multicolumn{5}{c}{+8.4\%} \\
\midrule
\textit{FakeMarrakesh}   & 0.979 & Simulation  & 2217 & 5.30 & 0.877 & 6.77 & 4.48 & 3799 & 9.41 & 0.351 & 11.64 & 15.97 \\
\textit{(Heron r2,}      &       & Hardware    & \textbf{2064} & \textbf{4.85} & \textbf{0.894} & \textbf{6.30} & \textbf{4.09} & 3920 & 9.73 & 0.309 & 12.01 & 16.06 \\
\textit{156 qubits)}     &       & Degradation & \multicolumn{5}{c}{$-$6.9\%} & \multicolumn{5}{c}{+3.2\%} \\
\bottomrule
\end{tabular}
\end{table*}

The final evaluation examines whether a readout trained on ideal reservoir states can still operate when the quantum reservoir is executed under realistic backend noise. This setting is important because the proposed framework freezes the quantum reservoir and performs inference without retraining. The trained FP32 readout is therefore applied directly to reservoir states generated under the \textit{IBM FakeTorino} and \textit{IBM FakeMarrakesh} noise models using 1024 shots per circuit. The simulation baseline is computed with the same shot count to isolate the effect of backend noise.
\begin{figure}
    \centering
    \includegraphics[width=1\linewidth]{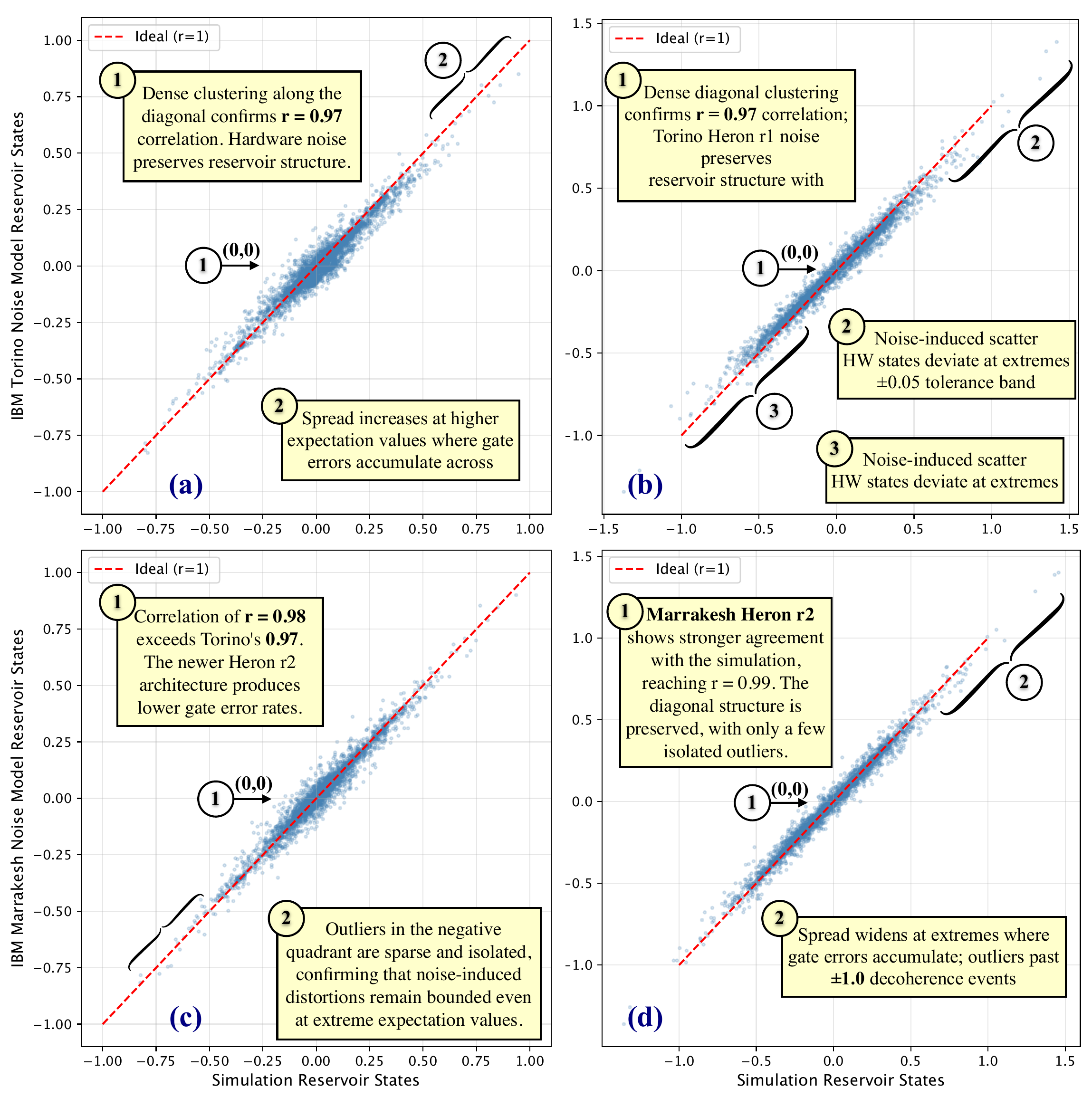}
    \caption{Simulation-versus-hardware reservoir states for the Tetouan and Spain datasets over 50 test samples under the \textit{IBM FakeTorino} and \textit{IBM FakeMarrakesh} noise models. Panels (a) and (c) show Tetouan, while panels (b) and (d) show Spain. In all cases, the points cluster tightly around the diagonal, showing that realistic hardware noise preserves the reservoir-state geometry. For Tetouan, the Pearson correlation increases from 0.97 on \textit{Torino} to 0.98 on \textit{Marrakesh}, indicating slightly better agreement with simulation under the newer \textit{Heron r2} backend. For Spain, both hardware models maintain a near-perfect correlation of 0.99 with low error, confirming that the reservoir representation remains stable under realistic device noise.}
    \label{fig:scatter}
\end{figure}
The hardware-noise results show different dataset-level responses. On Tetouan, both backends improve over their matched simulation baselines, with RMSE decreasing from 2172 to 2110~kWh on \textit{FakeTorino} and from 2217 to 2064~kWh on \textit{FakeMarrakesh}. This corresponds to negative degradation values of $-2.9\%$ and $-6.9\%$, suggesting a mild noise-regularization effect similar to the behavior reported in MTS-QRC~\cite{hamhoum2025multivariate}. On Spain, hardware noise increases the RMSE, but the degradation remains limited: $+8.4\%$ on \textit{FakeTorino} and $+3.2\%$ on \textit{FakeMarrakesh}. Across both datasets, \textit{FakeMarrakesh} gives the better hardware-noise result, indicating stronger agreement with the simulation baseline.

Fig.~\ref{fig:scatter} explains this behavior at the reservoir-state level. The noisy reservoir outputs remain closely aligned with the noiseless simulation outputs, with correlations ranging from $0.97$ to $0.99$. This indicates that backend noise perturbs the reservoir features but does not destroy the geometry needed by the trained readout. 

\begin{figure}
    \centering
    \includegraphics[width=1\linewidth]{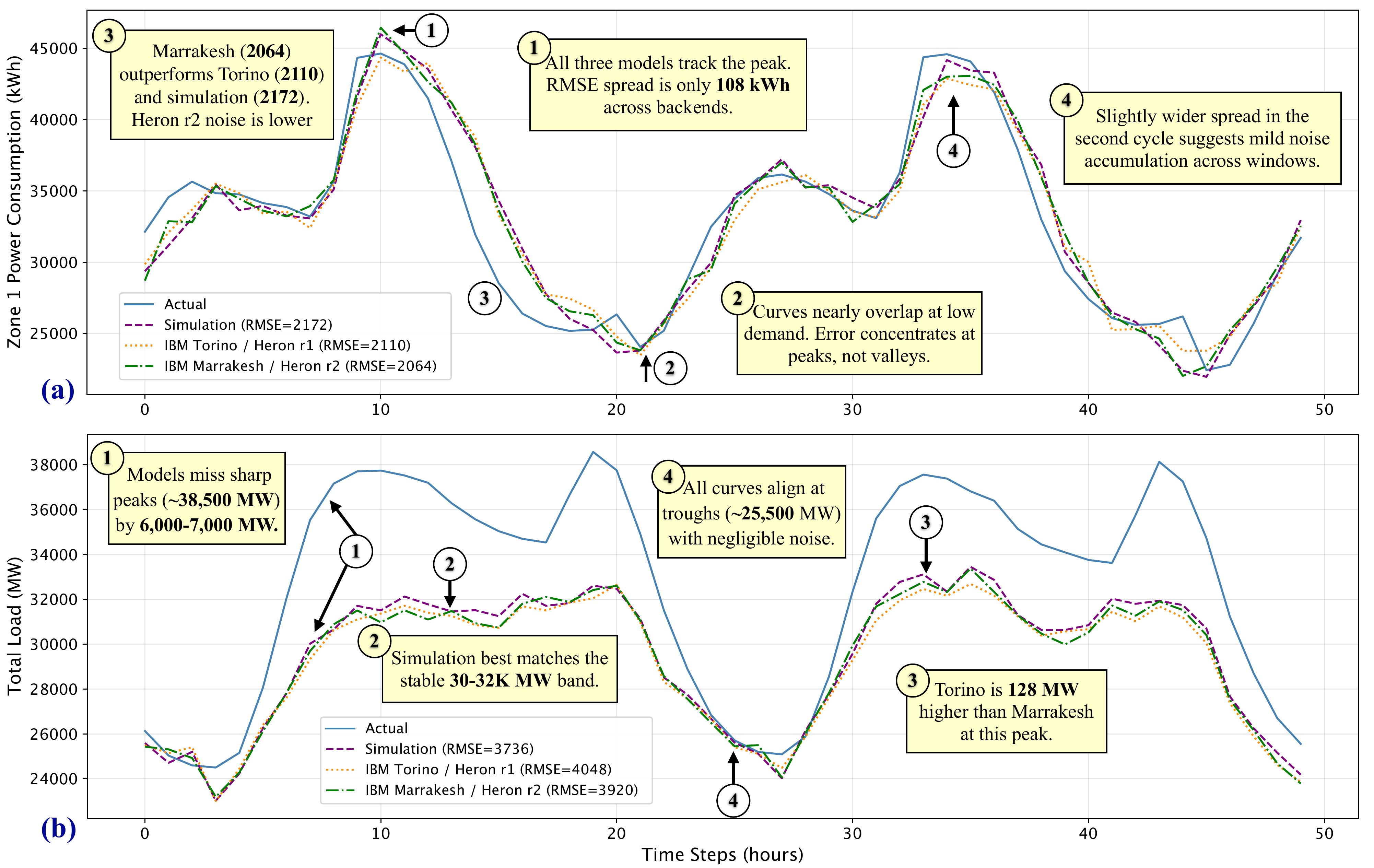}
    \caption{Prediction-level comparison between noiseless simulation and hardware-noise inference on 50 test samples. Panel (a) shows the Tetouan dataset, where both hardware backends match or slightly improve upon the simulation baseline, with RMSE decreasing from 2172~kWh in simulation to 2110~kWh on \textit{IBM Torino (Heron r1)} and 2064~kWh on \textit{IBM Marrakesh (Heron r2)}. Panel (b) shows the Spain dataset, where the simulation and hardware-based predictions follow the same hourly trend, with RMSE values of 3736, 4048, and 3920~MW for simulation, \textit{Torino}, and \textit{Marrakesh}, respectively. In both datasets, the hardware predictions remain close to the simulation baseline, indicating that realistic backend noise has a limited effect on forecasting quality.}
    \label{fig:hardware_predictions}
\end{figure}

The prediction-level comparison in Fig.~\ref{fig:hardware_predictions} shows that the main errors occur around sharp load peaks rather than across the full temporal profile. This suggests that the remaining gap is mainly linked to the smoothing behavior of the linear readout, not to a collapse of the reservoir under hardware noise. Overall, the hardware experiments show that the learned readout can transfer from ideal reservoir states to noisy reservoir states without retraining.

\subsection{Comparison with Related Work}
\begin{table*}[htpb]
\centering
\caption{Structural comparison of the proposed framework with related works.}
\label{tab:structural_comparison}
\footnotesize
\setlength{\tabcolsep}{3.5pt}
\begin{tabular}{lccccccc}
\toprule
\textbf{Method} & \textbf{Paradigm} & \textbf{Quantum} & \textbf{Trainable} & \textbf{Readout} & \textbf{PTQ} & \textbf{HW Noise} & \textbf{Application} \\
 & & \textbf{Training} & \textbf{Q. Params} & & & \textbf{Validation} & \\
\midrule
Fuzzy+MLP~\cite{alsalem2025hybrid} & Classical ML & None & N/A & MLP & \ding{55} & N/A & Energy (Tetouan) \\
ESN~\cite{hausser2026echo} & Classical RC & None & N/A & Ridge & \ding{55} & N/A & General TS \\
iQTransformer~\cite{ranilla2025quantum} & Variational QML & Gradient & $O(10^2)$ & Hybrid & \ding{55} & \ding{55} & General TS \\
MTS-QRC~\cite{hamhoum2025multivariate} & Quantum RC & None & 0 & Ridge & \ding{55} & \ding{51} (Heron r2) & Lorenz, ENSO \\
\textbf{This work} & \textbf{Quantum RC} & \textbf{None} & \textbf{0} & \textbf{Elastic Net} & \ding{51} & \ding{51} \textbf{(r1 + r2)} & \textbf{Energy (2 datasets)} \\
\bottomrule
\end{tabular}
\end{table*}
The proposed framework differs from existing load forecasting and quantum time-series models in three main aspects: the quantum reservoir is fixed, the trainable component is limited to a classical Elastic Net readout, and the readout is evaluated under both post-training quantization and realistic hardware-noise models. Since prior studies do not evaluate the same Tetouan and Spain datasets under the same protocol, the comparison is organized structurally rather than as a direct numerical benchmark. Table~\ref{tab:structural_comparison} summarizes these differences.

Classical approaches provide useful forecasting baselines, but they do not address the deployment setting studied here. Fuzzy+MLP models have been evaluated on the Tetouan dataset with strong full-precision performance \cite{alsalem2025hybrid}, while ESN-based methods share the frozen-reservoir principle with QRC \cite{hausser2026echo}. However, these approaches do not study quantum feature generation, finite-shot inference, post-training quantization, or hardware-noise transfer.
Among quantum approaches, iQTransformer relies on trainable quantum parameters optimized through gradient-based learning \cite{ranilla2025quantum}. This differs from the proposed QRC design, where the quantum circuit remains fixed and only the classical readout is trained. MTS-QRC is closer in structure because it also uses a frozen quantum reservoir and validates behavior under hardware noise \cite{hamhoum2025multivariate}, but it does not examine low-bit readout quantization or energy forecasting across the two datasets considered here.

The proposed framework fills a distinct gap by combining a fixed quantum reservoir, Elastic Net readout, low-bit PTQ, finite-shot evaluation, and hardware-noise validation on two energy load forecasting datasets. This comparison positions the work as a resource-aware and hardware-aware QRC study rather than a direct accuracy benchmark against models trained under different settings.

\subsection{Discussion and Key Insights}
The results show that the proposed QRC framework is tolerant to readout quantization up to a practical limit. Across both Tetouan and Spain, 6-bit PTQ preserves forecasting performance while reducing readout memory by 81.2\%. Below this point, degradation becomes dataset dependent: Spain declines gradually, while Tetouan is more sensitive under finite-shot simulation. This suggests that the readout can absorb moderate rounding errors, but very low precision can distort the learned forecasting function.
The consistent 6-bit operating point is important because the two datasets select different reservoir configurations. Tetouan uses a deeper Chebyshev-encoded reservoir with stronger coupling, while Spain uses a shallower $R_Z$--$R_Y$ encoded reservoir with weaker coupling. Despite these differences, both remain stable at 6-bit precision, indicating that the observed quantization resilience is not tied to one specific encoding or circuit depth.

The hardware-noise results show that the learned readout transfers from ideal reservoir states to noisy reservoir states without retraining. High simulation-to-hardware correlations indicate that realistic noise perturbs the reservoir features without destroying their main geometry. The remaining errors occur mainly around sharp demand peaks, suggesting that the main bottleneck is the smoothing tendency of the linear readout rather than quantization or hardware noise. Future work can therefore explore peak-aware losses, regime-specific readouts, or adaptive readout calibration while preserving the fixed quantum reservoir and low-precision deployment pipeline.

\section{Conclusion}\label{sec:conclusion}

This work presents a hardware efficient QRC framework for short-term load forecasting under the memory and precision constraints of edge deployed energy systems. The framework keeps the quantum reservoir entirely fixed, trains only a regularized linear readout on temporally aggregated reservoir states, and applies post training fixed point quantization to the readout weights at bit widths between 32 and 2-bits. Evaluated on the Tetouan and Spain datasets across noiseless and 512 shot regimes, the readout preserves full precision forecasting accuracy down to 6-bits while removing 81.2\% of its memory footprint, and the trained model transfers without retraining to the \textit{IBM FakeTorino} and \textit{FakeMarrakesh} noise backends, where \textit{Heron r2} even improves RMSE on Tetouan by 6.9\% over the matched simulation baseline. These results establish quantized QRC as a practical route toward deploying quantum forecasting on resource-constrained grid infrastructure, and motivate future work on peak aware readout designs, automated bit width selection, and end-to-end evaluation on physical quantum processors as the hardware matures.

\section*{Acknowledgment}
 This work was supported in part by the NYUAD Center for Quantum and Topological Systems (CQTS), funded by Tamkeen under the NYUAD Research Institute grant CG008.
\bibliographystyle{IEEEtran}
\bibliography{refs}

\end{document}